\documentclass[
reprint, 
twocolumn, 
secnumarabic,
tightenlines,
amssymb,
amsmath,
nobibnotes,
nofootinbib, 
floatfix,
aps, prd, 
showpacs,showkeys
]{revtex4-2}
\usepackage{hyperref}
\usepackage{multirow}%
\expandafter \ifx \csname package@font\endcsname\relax\else
\expandafter \expandafter \expandafter
\usepackage
\expandafter \expandafter
\expandafter{\csname package@font\endcsname}%
\fi
\usepackage{ulem} 
\usepackage{textcomp}
\usepackage{gensymb}

\newcommand{\n}[1]{\ensuremath{|\mathbf{#1}|}}

\usepackage{forest}

\newcommand{\etal}{\textit{et al.}}

\usepackage{cleveref}
\usepackage{graphicx} %

\begin{document}

\begin{abstract}
Employing the neural network framework, we obtain empirical fits to the electron-scattering cross sections for carbon over a broad kinematic region, extending from the quasielastic peak through resonance excitation to the onset of deep-inelastic scattering. We consider two different methods of obtaining such model-independent parametrizations and the corresponding uncertainties: based on the bootstrap approach and the Monte Carlo dropout approach. 
In our analysis, the $\chi^2$ defines the loss function, including point-to-point and normalization uncertainties for each independent set of measurements.  Our statistical approaches lead to fits of comparable quality and similar uncertainties of the order of $7$\%. 
To test these models, we compare their predictions to test datasets excluded from the training process and theoretical predictions obtained within the spectral function approach. The predictions of both models agree with experimental measurements and theoretical calculations. We also perform a comparison to a dataset lying beyond the covered kinematic region, and find that the bootstrap approach shows better interpolation and extrapolation abilities than the one based on the dropout algorithm.

\end{abstract}

\title{Empirical fits to inclusive electron-carbon scattering data obtained by deep-learning methods}

\author{Beata E. Kowal}
\email{beata.kowal@uwr.edu.pl}

\author{Krzysztof M. Graczyk}
\email{krzysztof.graczyk@uwr.edu.pl}

\author{Artur M. Ankowski}
\author{\\Rwik Dharmapal Banerjee}
\author{Hemant Prasad}
\author{Jan T. Sobczyk}

\affiliation{Institute of Theoretical Physics, University of Wroc\l aw, plac Maxa Borna 9,
50-204, Wroc\l aw, Poland}

\date{\today}%

\maketitle

\section{Introduction}

The stringent precision requirements of the long-base{\-}line neutrino-oscillation program create an urgent need to significantly improve the accuracy of available estimates of the cross sections for neutrino interactions with atomic nuclei~\cite{DUNE:2020lwj,Hyper-KamiokandeProto-:2015xww}. The complexity of this problem stems from multiple factors: the process of scattering involves multicomponent vector and axial currents, different interaction mechanisms can produce the same final states, experimental data with the desired precision are only available for nuclear cross sections, and flux-averaging greatly diminishes differences between different interaction channels.

A widely recognized way to address those difficulties is to leverage the similarities of electron and neutrino interactions~\cite{Ankowski:2022thw}. In electron scattering, only electromagnetic currents contribute, cross sections are higher by several orders of magnitude, beams are monoenergetic and adjustable, and precise data can be collected not only for complex nuclei, but also for deuterium and hydrogen. These advantages allow nuclear effects to be studied in detail using electron-scattering data.

Provided that the cross sections for neutrinos and electrons are calculated consistently and are sufficiently constrained by precise electron data, the only significant source of uncertainties in the neutrino predictions would be related to the axial contributions, to be determined from the near-detector measurements, possibly making use of the guidance provided by the lattice quantum chromodynamics computations.

In parallel to the efforts of developing models of nuclear response, it would be useful to explore novel techniques for obtaining model-independent estimates of nuclear cross sections. Deep neural networks (DNNs) are ideally equipped to provide such predictions. Indeed, DNNs are known for their excellent adaptive abilities~\cite{Cybenko_Theorem}, which enabled obtaining unbiased parametrizations of nuclear cross sections \cite{AlHammal:2023svo}, parton distribution functions~\cite{Forte:2002fg,DelDebbio:2004xtd}, vector \cite{Graczyk:2010gw,Graczyk:2011kh,Graczyk:2014lba,Graczyk:2014coa} and axial~\cite{Alvarez-Ruso:2018rdx} nucleon form factors, deeply virtual Compton form factors \cite{Kumericki:2011rz}, and nuclear charge radii~\cite{Dong:2021aqg}. 

In this article, we explore methods of obtaining such parametrizations of electron-scattering cross sections, independent of nuclear model assumptions, and pay special attention to estimating their uncertainties. For an analysis of this type, the abundance of available experimental data is critical. 

We consider the carbon target, for which experimental cross sections have been collected in various laboratories worldwide over the last five decades and span the broadest kinematic region~\cite{Ankowski:2020qbe}. These features are of paramount importance to maximize the predictive abilities of the resulting DNN models in the broad kinematic region of relevance for neutrino-oscillation experiments.

Using DNNs to parametrize electron-scattering data, we obtain both the best fits and an estimate of the associated uncertainties. In this analysis, we consider two different methods. 

In the first one, we adopt an ensemble approach, bootstrap-like model \cite{Tibshirani_96,Breiman1996}, similar to the one developed by the Neural Network Parton Distribution (NNPDF) group~\cite{Forte:2002fg}. The experimental data are used to generate bootstrap datasets. For each of the datasets, a~DNN fit is obtained. The model's prediction is obtained by averaging all the fits, and the square root of the variance gives the uncertainty. 

In the second approach, we employ the Monte Carlo (MC) dropout technique~\cite{JMLR:v15:srivastava14a}, in which DNN units are temporarily removed at random to prevent overfitting. This procedure is repeatedly used during the training and inference stages, resulting in random responses of the DNN. These responses are then averaged over to obtain the DNN prediction, and their distribution provides the uncertainty estimate. 

Our analysis pays special attention to systematic uncertainties related to data normalization, as previously done in Refs.~\cite{Graczyk:2011kh,Graczyk:2014lba,Graczyk:2014coa}. We introduce them as a penalty term in the loss function.  We use a DNN architecture with ten hidden layers and address the vanishing gradient problem (VGP) by implementing batch normalization layers~\cite{ioffe2015batch}. Our models are validated by comparisons to the test dataset and with predictions of the spectral function approach~\cite{Ankowski:2014yfa}. Eventually, we compare the model's prediction with higher energetic data from Ref.~\cite{Gomez:1993ri}.

A similar analysis of electron-scattering data was recently performed by Al~Hammal \textit{et al.} \cite{AlHammal:2023svo}, who considered nuclear targets ranging from $^4$He to $^{59}$Ni. Their loss function did not account for experimental uncertainties and was given as a relative absolute error. However, an effort was made to estimate the uncertainties due to the model parameter variations. The authors of Ref.~\cite{AlHammal:2023svo} adopted a~network with $35$ hidden layers and relied on the residual mechanism for the VGP. To test their model, Al Hammal \textit{et al.} compared their results with the predictions of GiBUU \cite{Mosel:2018qmv} and the SuSAv2 approach~\cite{Amaro:2019zos}. 

The  DNN approaches do not make any assumptions on whether the interaction takes place at the nucleus or nucleon level and do not separate the longitudinal and transverse response functions. As a consequence, the  DNN approaches are conceptually different from the phenomenological approaches that were successfully developed in the literature, e.g, for electron scattering on the proton~\cite{Christy:2007ve}, deuteron~\cite{Bosted:2007xd}, and carbon and oxygen~\cite{Bodek:2022gli,Bodek:2023dsr} targets.

Our paper is organized as follows. In Sec.~\ref{Sec:Data}, we discuss the experimental data underlying our analysis and provide the rationale for the imposed kinematic cuts. In Sec~\ref{Sec:NN}, we introduce our statistical model, providing details of the loss function, the methods for estimating uncertainties, and the network architectures. In Sec.~\ref{Sec:results}, we report and analyze our numerical results. Finally, we state our conclusions in Sec.~\ref{sec:summary}.

\section{Electron-scattering data}
\label{Sec:Data}
For a DNN to successfully learn electron-scattering cross sections, it is essential to provide it with a sufficient amount of experimental data. In this analysis, we consider the carbon target, which has been studied extensively in past experiments; see Ref.~\cite{Ankowski:2020qbe} for a recent review.

\begin{figure}[!b]
\centering
\includegraphics[width=0.9\columnwidth]{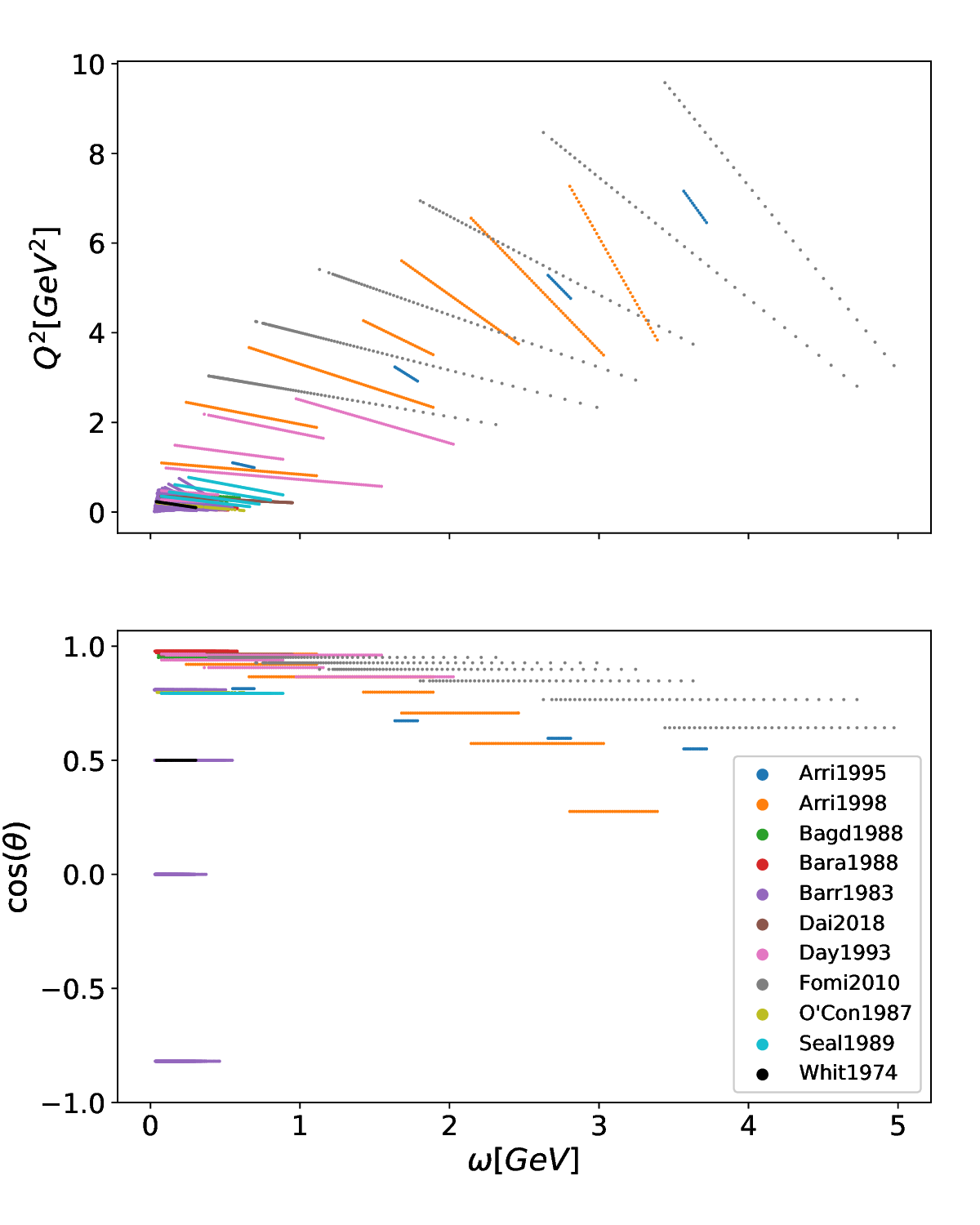}
\caption{Kinematic domain covered by the experimental data~\cite{Arrington:1995hs,Arrington:1998ps,Bagdasaryan:1988hp,Baran:1988tw,Barreau:1983ht,Dai:2018xhi,Day:1993md,Fomin:2010ei,O'Connell:1987ag,Sealock:1989nx,Whitney:1974hr} considered in our analysis, shown in the planes of (top) energy transfer and four-momentum transfer, and (bottom) energy transfer and cosine of the scattering angle.}
\label{plot_data}
\end{figure}

The available data~\cite{Arrington:1995hs,Arrington:1998ps,Bagdasaryan:1988hp,Baran:1988tw,Barreau:1983ht,Dai:2018xhi,Day:1993md,Fomin:2010ei,O'Connell:1987ag,Sealock:1989nx,Whitney:1974hr,Benhar:2006er} are in the form of double-differential cross sections ${d^2 \sigma}/{d\omega d\Omega}$  for inclusive electron scat{\-}ter{\-}ing---in which only the final kinematics of the electron is measured---at various beam energies $E$ and scattering angles $\theta$, given as a function of the energy $\omega$ that the interaction transferred to the nucleus.

We consider a broad kinematic region, in which different interaction mechanisms are known to play an important role, starting from quasielastic scattering at low energy transfers, to pion production through resonance excitation in the intermediate regime, and to the onset of deep-inelastic scattering at the highest $\omega$ values.

At very low values of energy transfer, elastic scattering off the whole nucleus may occur, and inelastic interactions with the nucleus may lead to an excitation of the giant dipole resonance or a discrete nuclear state. Due to the scarcity of data in this region, accounting for these processes is not currently feasible in our approach. We remove their contributions by applying a cut,
\begin{equation}
\label{Eq:omega_min}
\omega>\omega_\text{min} =\frac{  M_A E_x + 0.5 E_x^2 + E^2 (1 -\cos\theta) }{  M_A + E (1 - \cos\theta) },
\end{equation}
where $\omega_\text{min}$ is the energy transfer required to excite the nucleus of the mass $M_A$ with the energy $E_x$, and $\theta$ is the scattering angle. Typically, $\omega_\text{min}\approx E_x+t_A$, with $t_A$ being the nuclear recoil energy. In our analysis, the  $E_x$ value is set to $26$~MeV~\cite{Ankowski:2014yfa}.

We do not include the measurements reported by Gomez \etal~\cite{Gomez:1993ri}, who studied how the EMC effect depends on the nuclear mass number. Those seven points, each collected for a~different $(E, \cos\theta)$ pair, provide useful constraints on theoretical models, but they cannot be readily accommodated within our machine-learning approach. We make use of those data to investigate the extrapolation properties of our models.

The kinematic region spanned by the data used in our analysis is presented in Fig.~\ref{plot_data}, projected in the $(\omega, Q^2)$ and $(\omega, \cos\theta)$ planes, $Q^2$ being the four-momentum transfer. While at low $Q^2$ and at low $\theta$, there is an abundance of the experimental data; this is not the case in general. Even for carbon, the best-studied nuclear target, there are still large swaths with very scarce information on the cross sections. This issue is challenging to machine-learning models based on data alone.

\begin{table}[!tb]
\caption{Summary of the data used in this analysis. The numbers of points refer to the data surviving the cut~\eqref{Eq:omega_min}.}\label{tab:syserrors}
\begin{ruledtabular}
\begin{tabular}[t]{llrr}
\multirow{2}{*}{Reference} & \multirow{2}{*}{Abbrev.}  &  Norm. & Number \\
& & uncert. & of points\\
 \hline
Arrington \etal~\cite{Arrington:1995hs} & Arri1995 &  $4.0\%$  & 56 \\
Arrington \etal~\cite{Arrington:1998ps} & Arri1998 &  $4.0\% $  &  398 \\
Bagdasaryan \etal~\cite{Bagdasaryan:1988hp} & Bagd1988 & $ 10.0\%$  & 125 \\
Baran \etal~\cite{Baran:1988tw} & Bara1988 & $ 3.7 \%$  &  259 \\
Barreau \etal~\cite{Barreau:1983ht} & Barr1983 & $ 2.0\%$  &  1243 \\
Dai \etal~\cite{Dai:2018xhi} & Dai2018 & $ 2.2 \%$  &  177 \\
Day \etal~\cite{Day:1993md} & Day1993 & $3.4\%$  &  316 \\
Fomin \etal~\cite{Fomin:2010ei} & Fomi2010  & $ 4.0\%$  &  359 \\ 
O’Connell \etal~\cite{O'Connell:1987ag} & O’Con1987 & $ 5.0\%$  & 51 \\
Sealock \etal~\cite{ Sealock:1989nx} & Seal1989 & $ 2.5\%$  & 250 \\
Whitney \etal~\cite{Whitney:1974hr} & Whit1974 & $ 3.0\%$  & 31 \\
\hline
Total &  &  & 3265  
\end{tabular}
\end{ruledtabular}
\end{table}

In Table~\ref{tab:syserrors}, we summarize the experimental data and provide the normalization uncertainties (`Norm. uncert.') assumed in our analysis.

\section{Neural network framework}
\label{Sec:NN}

\subsection{Likelihood analysis}
\label{Sec:ML:Likelihood}

As discussed in the previous section, in this analysis, we consider $11$ independent  datasets. We denote the $k$th dataset containing $N_k$ points as 
\begin{equation}
 \mathcal{D}_k = \{  ( E_k^{i }, \theta_k^{i}, \omega_k^i ,d \sigma_{k}^{i} , \Delta d\sigma_{k}^{i}): \;\; i = 1,\dots, N_k\}, 
\end{equation}
where $d\sigma_{k}^{i}$ and $\Delta d\sigma_{k}^{i}$ are the $i$th  measurement in $k$th dataset and its corresponding uncertainty, $N_k$ number of data points in $k$th dataset. The full dataset reads $\mathcal{D} = \mathcal{D}_1 \cup \dots \cup \mathcal{D}_{11}$. The total uncertainty $\Delta d\sigma_{k}^{i}$ combines the statistical and systematic uncertainties, characteristic to a given experiment.

The global analyses of data for electron-proton scattering \cite{Arrington:2003df,Alberico:2008sz}, charged current single pion production \cite{Graczyk:2009qm}, and quasielastic neutrino-deuteron scattering \cite{Alvarez-Ruso:2018rdx} found that including normalization uncertainties is essential to obtain consistent fits.

In the present paper, we follow the same  philosophy. To fit the data, we introduce the following loss function~\cite{Graczyk:2011kh}:
\begin{equation}
\chi_\text{tot} = \sum_{k=1}^{11} \left[ \chi_k^2(\lambda_k) + \frac{1}{2}\left(\frac{1 - \lambda_k }{\Delta \lambda_k}\right)^2\right],
\end{equation}
where
\begin{equation}
\chi_k^2(\lambda_k) = \frac{1}{2} \sum_{i=1}^{N_k} 
\left(\frac{d \sigma_{k}^i -  \lambda_k d \sigma_i^\text{fit} (E_k^i, \theta_k^i)}{\Delta d \sigma_{k}^i}\right)^2,
\end{equation}
with $\Delta \lambda_k$ denoting the overall  systematic uncertainty for kth dataset, and $\lambda_k$ being a subject of optimization. In order to optimize the $\lambda_k$'s, we follow the algorithm proposed in Refs. \cite{Graczyk:2011kh,Graczyk:2014lba,Graczyk:2014coa}.

\begin{figure}
\centering{\includegraphics[width=\columnwidth]{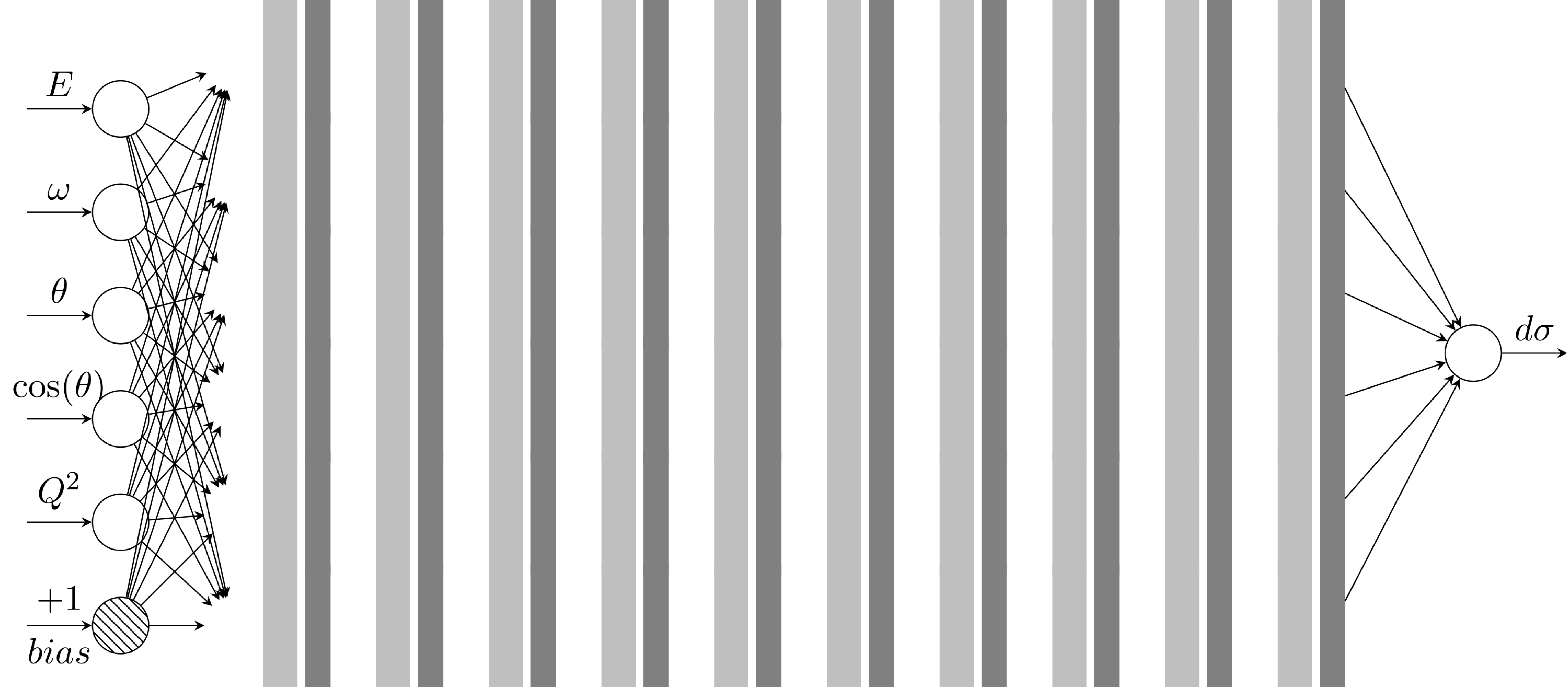}}
\caption{The neural network architecture for model A. The light gray box corresponds to the hidden layer of fully connected neurons with ReLU activation function. The dark gray box denotes the batch normalization layer. In the output, there is the sigmoid function. In the architecture for model B, not depicted, each fully connected layer is followed by the dropout layer. In both models A and B, the network has ten layers of hidden units, consisting of $300$ units each.}
\label{plot_nn}
\end{figure}

\subsection{Neural network models}

Our analysis aims to obtain a function that returns the value of differential cross section for electron-carbon scattering, given an input value of energy, scattering angle, and energy transfer. We use a DNN parametrization for this purpose.

While	 DNNs are known for their adaptive ability, they are also known for their limitations \cite{Bishop_book,franoischollet2017learning}. 	For example, when the input or output data values are too large (in absolute value), the network is unable to relate the input to the output, or the optimization algorithms work ineffectively.

In our problem, the output is the cross section, which ranges from $10^{-5}$ to
$10^{6}$ nb/(sr GeV). Therefore, to perform the optimization, we rescale the output as
\begin{equation}
    d\sigma \rightarrow  \left(\frac{10^9}{137^2 E \cos(\theta/2) }\frac{\cos^2(\theta/2) }{4 E^2 \sin^4(\theta/2)} \right)^{-1} d\sigma, 
    \label{scaling}
\end{equation}
where the beam energy, $E$, is given in GeV and the double differential cross section, $d\sigma$, in nb/(sr GeV).

\begin{figure}
\includegraphics[width=0.99\columnwidth]{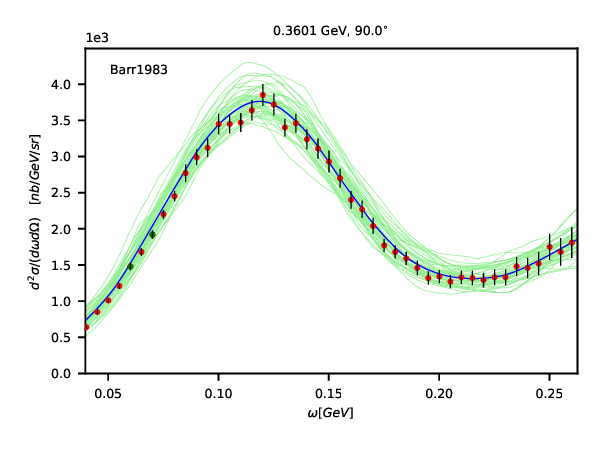}
\caption{Illustration of the algorithm used in model A to estimate the fit's uncertainty. Bootstrap datasets are generated from the original data according to the normal distribution (not depicted), and fits are performed $50$ times (green lines). The cross section's estimate corresponds to the mean of the fits (black line), and the uncertainty corresponds to their standard deviation (not shown). The data are taken from Ref.~\cite{Barreau:1983ht}.}
\label{single_plot}
\end{figure}

To accelerate the optimization process and help DNN capture the proper normalization, we extend the number of input variables from three to five $(E, \omega, \theta, \cos\theta, Q^2)$. The last two are the functions of the first three, but keeping such input accelerates the optimization. The same procedure was employed by NNPDF group \cite{Forte:2002fg} to fit parton distribution functions, and by Al Hammal \textit{et al.}~\cite{AlHammal:2023svo} to fit electron-nucleus scattering data.

We consider two types of DNN architecture. The first one, statistical model A, consists of the input layer, followed by ten blocks and an output layer. The network structure is shown in Fig.~\ref{plot_nn}. The input is five-dimensional, and the output is one-dimensional. Each block has a hidden layer with $300$  units fully connected with the previous module and batch-normalization layer. We keep ReLU activation functions in each module. The output activation function is given by a sigmoid function. The batch-normalization layer \cite{ioffe2015batch} is essential to obtain successful fits. This type of layer helps to maintain a proper normalization of the outputs that, as a result,  prevents the gradients from vanishing and the optimization process from stopping, which is a problem typical for DNNs.  Additionally, the batch-normalization layer naturally regularizes the model parameters. Hence, such a model should not tend to overfit the data.  

The second network, statistical model B, has a similar structure, but each module consists of a fully connected layer followed by a batch-normalization layer and the dropout layer. The dropout layer was introduced by Srivastava \textit{et al.} \cite{JMLR:v15:srivastava14a} to regularize the network parameters, prevent the model from overfitting and, as a consequence, increase the predictive abilities of the networks.


\subsection{Uncertainty estimation}

\subsubsection{Uncertainties in deep learning}

One of the challenges in modeling data using DNNs is obtaining systems with good generalization abilities. Many adaptive units define the DNNs and can easily overfit the data, resulting in poor generalization. Another challenge is estimating how uncertain the networks are in their predictions.

Evaluation of uncertainties in deep neural network predictions is difficult. DNNs are defined by many parameters, and standard statistical techniques either do not work or are inefficient. There are two primary sources of (predictive) uncertainty: epistemic (model's uncertainties due to its parameter dependence and structure) and aleatoric (data uncertainty) \cite{H_llermeier_2021}.

In recent years, an effort has been made to develop methods for estimating the predictive uncertainties \cite{survey_uncertainty_in_dnn}. However, an optimal method has not been established yet, as each of the developed techniques is known to have some limitations and problems.

There are two classes of the most
popular methodologies for estimating uncertainties~\cite{survey_uncertainty_in_dnn,ABDAR2021243,smith2018understanding}: the ensemble and Bayesian methods.
In the first one, some number (from dozens to thousands)
of the networks are trained. The model's prediction
is given by the mean of the model’s responses, and the uncertainty is determined from their standard deviation. The second class of methods has roots in Bayesian statistics, which offers tools that capture both epistemic and aleatoric uncertainties.

An example of an ensemble method is the bootstrap approach, which originates from frequentist
statistics \cite{Efron_bootstrap} and can be considered in nonparametric
and parametric forms. In the first case, the idea is to create $B$ bootstrap sets by sampling with repetition. In the other, the data are drawn from the generating probability distribution. It is crucial to notice that the bootstrapping leads to the \textit{``poor man's Bayes posterior``} \cite{10.1214/12-AOAS571}. Hence, one should expect that bootstrap and Bayesian approaches lead to similar density distributions.

The bootstrap technique for neural networks was adapted independently by Tibshirani \cite{Tibshirani_96} and Breiman \cite{Breiman1996}. This method is known in the literature as bootstrap aggregation or bagging; see Refs.~\cite{Breiman1996,Bishop_book}, and it 
is commonly used in the neural network community. One reason for its popularity is its low computational requirements compared to Bayesian techniques. In fact, the recommended number of neural models trained on the bootstrap datasets varies from $25$ to $200$. 

The bootstrap approach estimates both types of predictive uncertainties due to the uncertainties in measurements and the variation of network parameters. Moreover,  as we remarked above, bootstrapping leads to a similar ''posterior'' distribution as sampling in parameter space in Bayesian Monte Carlo approaches. In the first, the prior probabilities are modeled by the generated bootstrap datasets; in the other, they are postulated based on objective Bayesian assumptions.

The Bayesian methods are considered state-of-the-art for estimating uncertainties. The main goal is to evaluate the set of posterior density probabilities. The Bayesian approach allows us to predict both aleatoric	 and epistemic uncertainties. Moreover, within the Bayesian approach, it is possible to compare the various neural network models. The main difficulty is evaluating the posterior densities.

One popular method with roots in Bayesian statistics but is simple in implementation is the Monte Carlo dropout technique~\cite{gal2015dropout}. It is based on the observation that data analysis using a one-hidden neural network with an infinite number of neurons corresponds to the Gaussian process~\cite{Neal:1995}  --- one of the well-formulated methods of Bayesian statistics~\cite{Rasmussen_Willimas_book}. A single-fit network model is obtained as the result of the MC dropout.

\subsubsection{Models A and B}

We consider two models to estimate the uncertainties of the network predictions. In model A, we adapt the bootstrap method as Refs.~\cite{Forte:2002fg,DelDebbio:2004xtd}.  Its idea is to use experimental measurements to produce the so-called bootstrap datasets and fit every set by the corresponding network model. The number of data points in the bootstrap dataset is the same as in the original one. A bootstrap point is obtained by sampling from a normal distribution. If the original data point is characterized by central value $d \sigma_k^i$ and uncertainty $\Delta d \sigma_k^i$, the  bootstrap datum reads
\begin{equation}
d \sigma_k^\text{clone, $i$} = d \sigma_k^i + r \Delta d \sigma_k^i, 
\end{equation}
where $r$ is drawn from the standard normal distribution.   We treat the normalization uncertainties as described in the previous sections and introduce $\chi^2$ as a loss function to optimize the network's parameters.

We split the full dataset into the training and test datasets in proportions $9$ to $1$. For the training data, $50$  bootstrap datasets are created. Then, we obtain the network fit for every  bootstrap dataset. The mean gives the model's response over predictions of all $50$ fits, while the corresponding square root of variance  defines uncertainty. In Fig.~\ref{single_plot} we illustrate how the algorithm works. The figure shows the experimental data together with  $50$ fits to the  bootstrap datasets.

In model B, we use the Monte Carlo dropout technique \cite{JMLR:v15:srivastava14a}.  Its idea is to keep the dropout layer after every  layer of units. This layer switches off each unit in the network with user-defined probability $p$. Initially, this method was introduced to regularize the training process to prevent overfitting and improve the model's generalization ability. However, when one keeps the dropout layer active in the inference mode, the network's prediction becomes random.  To obtain the network prediction, one computes the network’s response dozens of times (in our case,
50 times) and takes the average~\cite{gal2015dropout}. The uncertainty is given by the corresponding variance. 
In this method, the predictive uncertainty and the model's performance depend on the hyperparameter $p$, the dropout rate, which must be fine-tuned.

We will employ two criteria to fine-tune the value of $p$. The first criterion is to yield the model with the lowest generalization error on the test dataset. The second criterion is for the model to reproduce the results of model~A.  As discussed in Sec.~\ref{Sec:results}, we determine the dropout rate $p = 0.01$ to give the best MC dropout model.

\subsubsection{Overfitting}

Neural network models can have problems with generalization. If a model has a low capacity for describing analyzed data, it underfits it. On the other hand, when a model is too complex (that is, it has too many adaptive units), it may overfit the data. In both cases, the predictive abilities of the model are low. This problem is known as a bias-variance dilemma~\cite{Geman1992}. The idea is to find the optimal neural network that describes the analyzed data well and has good predictive abilities.

There are various methods developed to prevent models
from data overfitting. 
The simplest solution is to add a penalty term (regularization) to the loss, so that the algorithm does not reach the local minimum of the loss but converges
to the configuration of weights in the neighborhood of the minimum. 

In our analyses, we adopt that method. We optimize the model parameters using the AdamW algorithm with the weight decay of $0.004$, determined during the pre-analyses. Having non-zero weight decay corresponds to considering the loss of the form 
\begin{equation}
\chi^2_{\text{tot}}(\mathrm{mini}\; \mathrm{batch}) + \frac{\mathrm{weight}\; \mathrm{decay}}{2} \sum_{i} w_i^2,
\end{equation}
where $w_i$'s are the weights (parameters) of the network.

Analyzing model A, we average over an ensemble of models. Consequently, even if a single model overfits the bootstrap data, this is not the case for the average prediction.
In model B, the dropout layers regularize the model. Both analyses consider DNNs with batch normalization layers, which also naturally regularize the model. 

Finally, we would like to mention that there is a significant difference between the generalization
abilities of the small models (low number of adaptive units), such as one-hidden layer neural networks, and DNNs employed in this article. When the number of data constraints (in our case, data
points) is much smaller than the number of model parameters, the DNNs naturally tend to generalize well, even without regularization procedures ~\cite{zhang2017understanding}. Here, the neural network is defined by more than
$800,000$ parameters, while the number of data points is
only about $3,000$.

\begin{figure}
\includegraphics[trim={25 50 40 50},clip,width=0.90\columnwidth]{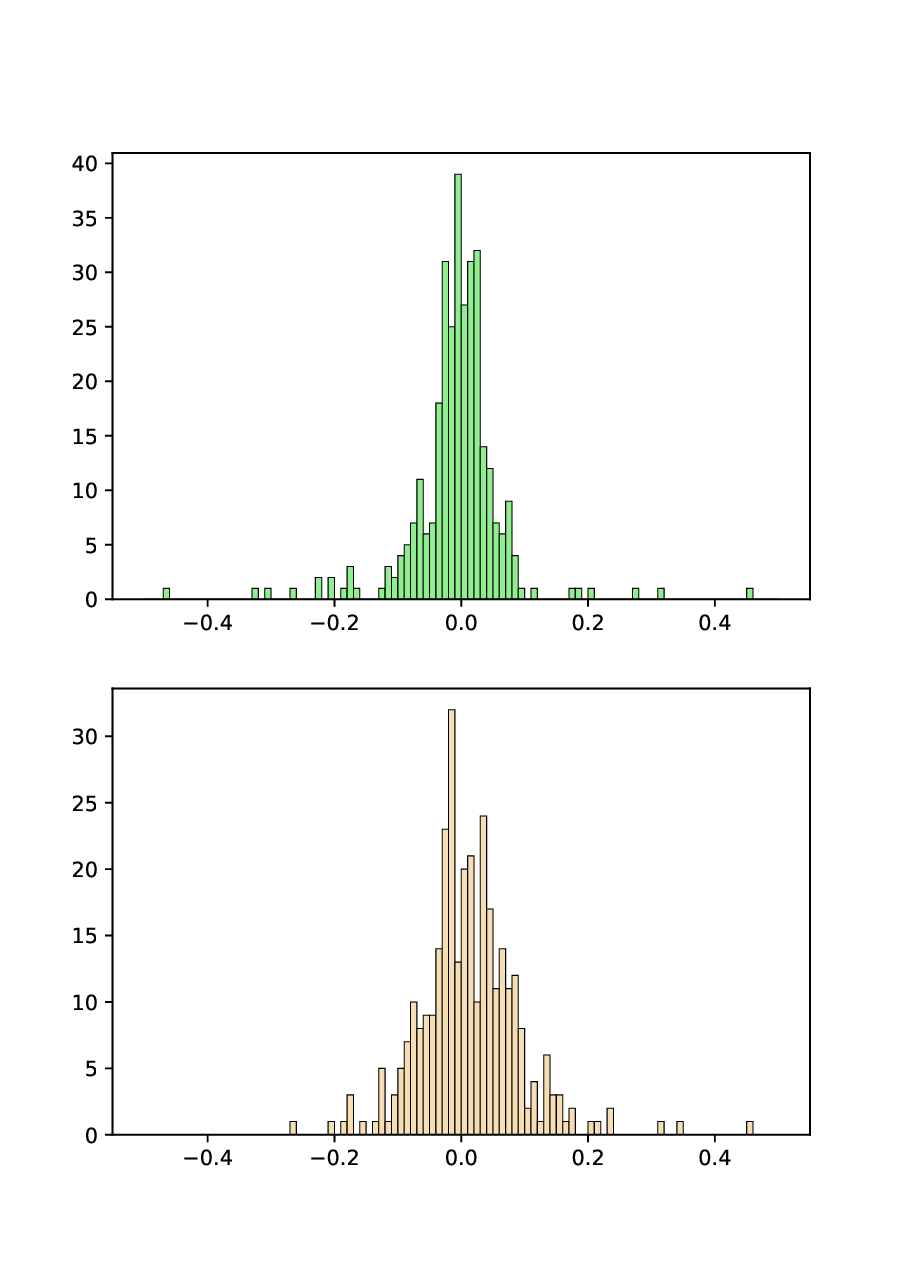}
\caption{Histograms of the normalized residuals, $(d\sigma_i - d \sigma_i^\text{fit})/d\sigma_i$, for (top) model A and (bottom) model B. Test data are included.}
\label{Fig:histogram_rel_error}
\end{figure}
\begin{figure}
\includegraphics[trim={25 50 40 50},clip,width=0.90\columnwidth]{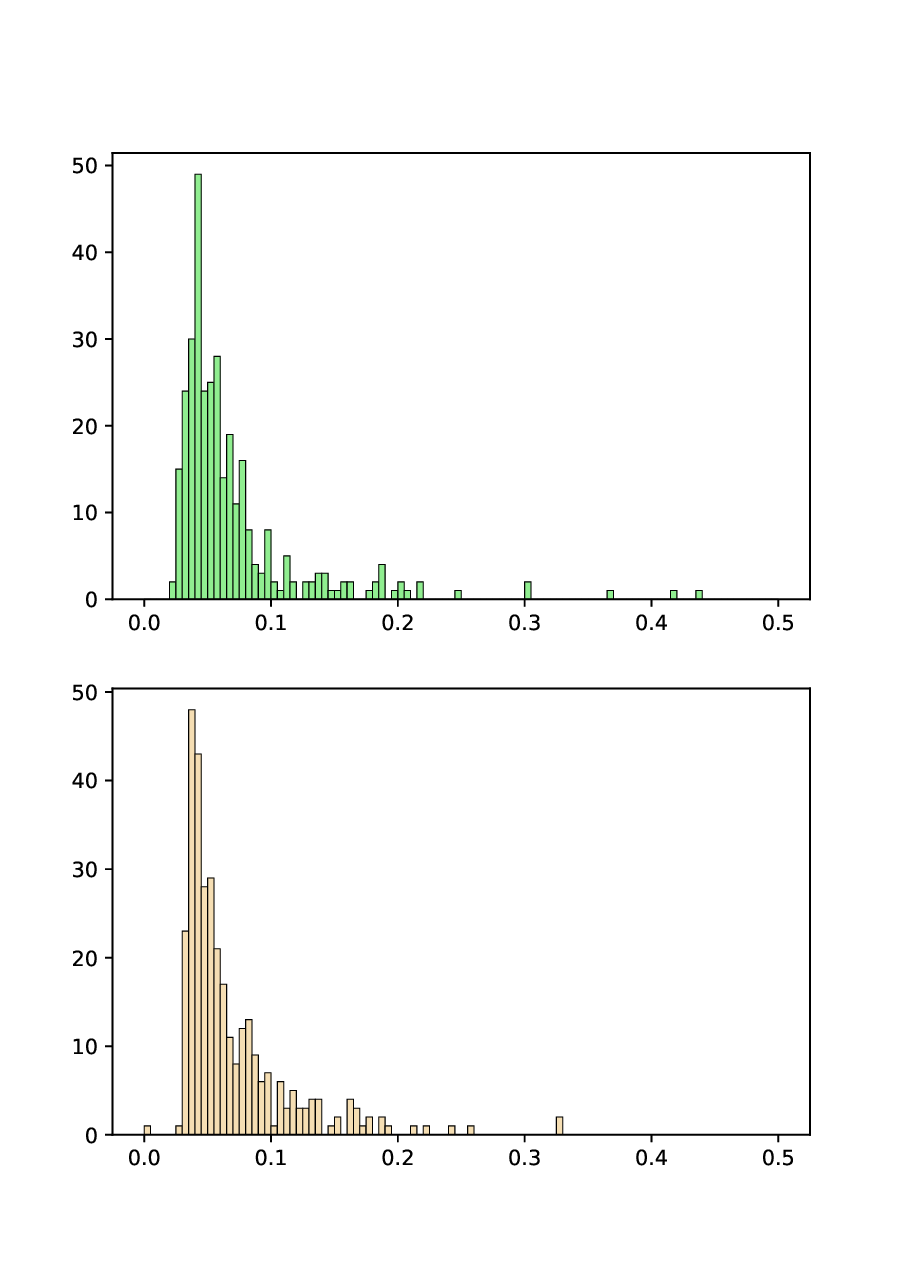}
\caption{Histograms of the standard deviation normalized by the experimental central value of (top) model A and (bottom) model B. Test data are included.}
\label{Fig:histogram_stnd_dev}
\end{figure}

\begin{figure*}
\includegraphics[width=0.9\textwidth]{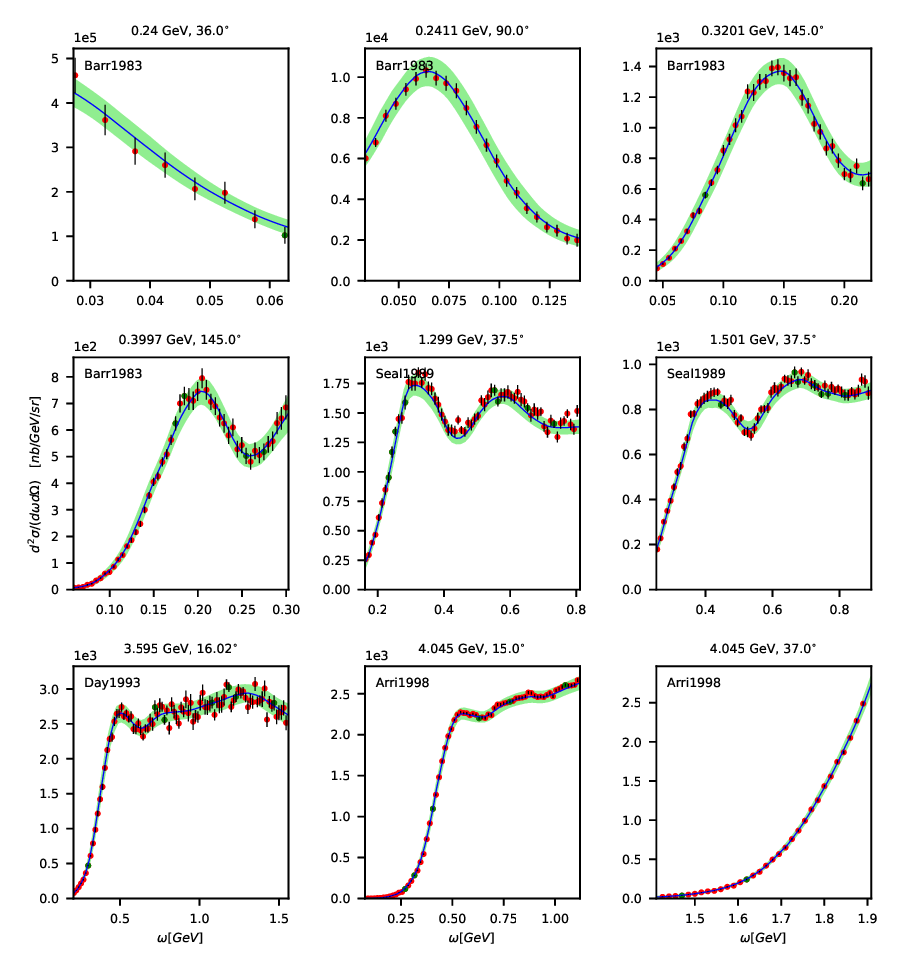}
\caption{Double-differential cross section ${d^2 \sigma}/{d\omega d\Omega}$ for inclusive electron scattering on carbon. We compare the predictions of model A to the experimental data from Refs.~\cite{Barreau:1983ht,Sealock:1989nx,Day:1993md,Arrington:1998ps}. The shaded areas denote the $1\sigma$ uncertainties. The panels are labeled with the beam energy and scattering angle values. The red (blue) points represent the training (test) dataset. The predictions are not rescaled according to the determined normalization parameters.}
\label{Fig:Dropout3x3_clone}
\end{figure*}
\begin{figure*}
\includegraphics[width=0.9\textwidth]{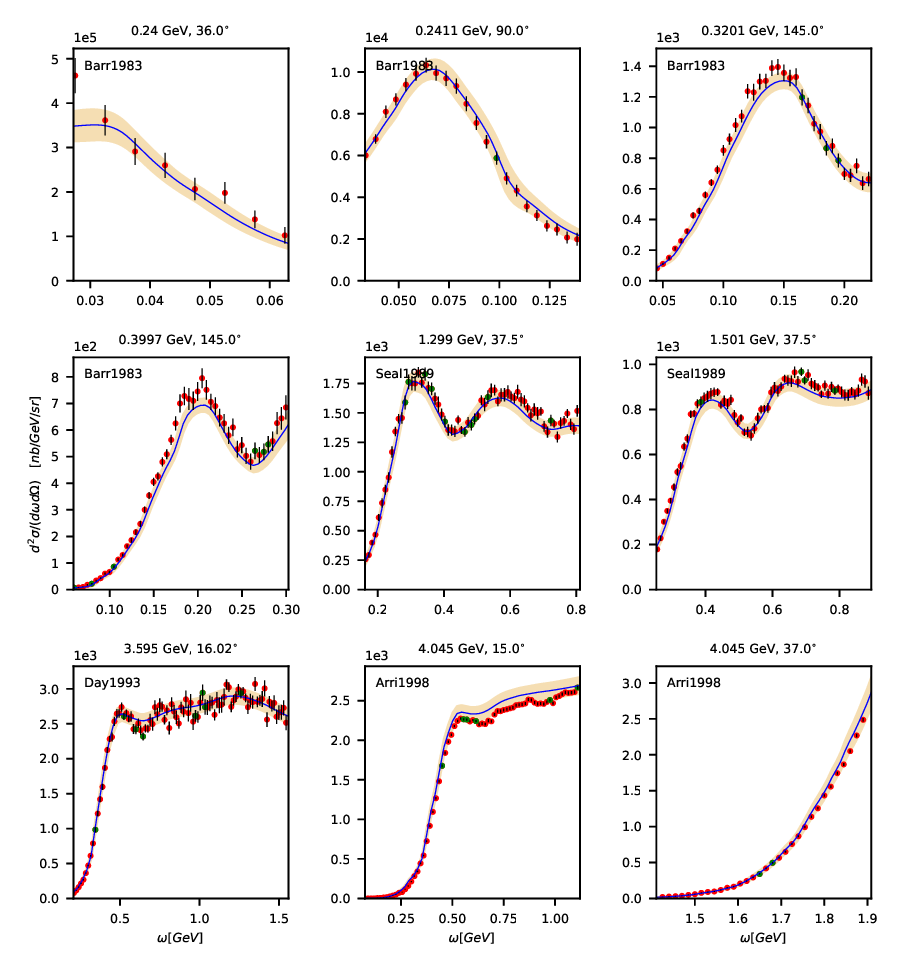}
\caption{Same as Fig.~\ref{Fig:Dropout3x3_clone} but for model B.}
\label{Fig:Dropout3x3_dropout}
\end{figure*}

\begin{table}
\caption{Values of the normalization parameters $\lambda$ obtained in models A and B.
\label{tab:syserrors-results}
}
\begin{ruledtabular}
\begin{tabular}[t]{llcc}
\multirow{2}{*}{Abbrev.}   &  Norm. & \multirow{2}{*}{model A} & \multirow{2}{*}{model B} \\
&  uncert. &  & \\
 \hline 
 Arri1995 &  $4.0\%$  & 1.01 &  1.02\\ 
 Arri1998 &  $4.0\% $  & 1.00  & 0.96\\
 Bagd1988 &  $10.0\%$  & 1.03 & 1.06\\
 Bara1988 & $ 3.7 \%$  & 1.01 & 0.98\\
 Barr1983 & $ 2.0\%$   & 0.99 & 1.02\\
 Dai2018 & $ 2.2 \%$   & 1.00 & 0.97\\
 Day1993 & $3.4\%$     & 0.99 & 0.98\\
 Fomi2010  & $ 4.0\%$  & 1.01  & 0.96\\ 
 O’Con1987 & $ 5.0\%$   & 1.02  & 1.01\\
 Seal1989 & $ 2.5\%$  & 1.02 & 1.04\\
 Whit1974 & $ 3.0\%$  & 0.93  & 0.93\\
 
\end{tabular}
\end{ruledtabular}
\end{table}

\section{Numerical results}
\label{Sec:results}

We performed both analyses employing the Jax package~\cite{jax2018github} and cross-checked the obtained results by using the Keras package \cite{chollet2015keras}. 
We ran optimization for no more than about 60,000 epochs. To optimize the models, we used the AdamW \cite{loshchilov2019decoupled} algorithm~\cite{loshchilov2019decoupled}. The models were regularized by keeping the value of weight decay as $0.004$. The training was done in the minibatch configuration with five batches. 

As discussed in the previous section, model~B must be calibrated. We considered the network with several values of dropout rate $p$. The model with $p=0.01$ was characterized by the lowest value of $\chi^2$ computed on the test dataset, see Table \ref{Table_ModelB_pvalues}.
\begin{table}
\caption{Generalization error on the test dataset for model~B, with the total number of data points $N$. \label{Table_ModelB_pvalues}}
\begin{ruledtabular}
\begin{tabular}{ c|l|l|l|l|l|l|l|l }
$p$  & 0.005 & 0.01 & 0.02 & 0.03 & 0.04 & 0.10 & 0.16 & 0.20 \\ 
\hline
$\chi^2/N$(test)  & 1.60  & 1.15 & 2.05 & 2.32 & 2.79 & 2.91 & 3.77 & 5.23 \\
\end{tabular}
\end{ruledtabular}
\end{table}

We also found that when $p = 0.01$ in model B, the obtained distributions of normalized residuals and normalized standard deviations were similar to those for model A. As an example, in Figs.~\ref{Fig:histogram_rel_error}~and~\ref{Fig:histogram_stnd_dev}, we show the histograms computed for the test dataset. Additionally, in Table \ref{Table_std_mean}, we present the mean of normalized standard deviations obtained for model~B with various $p$ rates. In the considered range, the lower the value of $p$, the lower the mean uncertainty.  As the mean in model~A is about $7\%$, the consistency requirement favors the value $p = 0.01$ in model~B.

\begin{table*}
\caption{The mean of normalized standard deviations (Norm. uncert.) obtained for model~B with various $p$ rates \label{Table_std_mean}}
\begin{ruledtabular}
\begin{tabular}{ c|l|l|l|l|l|l|l|l }
$p$  & 0.005 & 0.01 & 0.02 & 0.03 & 0.04 & 0.10 & 0.16 & 0.20 \\ 
\hline
Mean (Norm. uncert.) & 0.05  & 0.07 & 0.08 & 0.10 & 0.12 & 0.14 & 0.21 & 0.27 \\
\end{tabular}
\end{ruledtabular}
\end{table*}
As both the $\chi^2$ and model-consistency arguments support the dropout rate $p = 0.01$, we employ this value in our calculations with model~B.

In Figs.~\ref{Fig:Dropout3x3_clone} and \ref{Fig:Dropout3x3_dropout}, we present comparisons of the predictions of models A and B to six datasets~\cite{Barreau:1983ht,Sealock:1989nx,Day:1993md,Arrington:1998ps}. The data span  a broad kinematic region, corresponding to energy transfers between $\approx$30 MeV and $\approx$1.9 GeV, and momentum transfers $\n q$ between $\approx$0.14 and $\approx$2.7 GeV. 

The first six panels of Figs.~\ref{Fig:Dropout3x3_clone} and \ref{Fig:Dropout3x3_dropout} sample the $\n q$ values below 1 GeV, where the cross section is dominated by quasielastic scattering and the $\Delta$-resonance excitation, and nuclear effects are the most pronounced. Model A reproduces the data with excellent accuracy, both at low and high scattering angles, although contributions of the mechanism of interactions involving two-body currents are expected to be very different in these regimes. It outperforms model B, for which the agreement is very good nevertheless.

The last three panels of Figs.~\ref{Fig:Dropout3x3_clone} and \ref{Fig:Dropout3x3_dropout} probe the transition of the cross section from the region where excitation of higher resonances plays an important role to where it can be described in terms of deep-inelastic scattering. Despite of the complexity of the interaction mechanisms at these kinematics, both models are able to reproduce the data with remarkable accuracy.  It is important to keep in mind that the normalization parameters are \textit{not} included in the results
presented in  Figs.~\ref{Fig:Dropout3x3_clone} and \ref{Fig:Dropout3x3_dropout}.

As discussed in Sec.~\ref{Sec:NN}, the normalization parameters are incorporated in our fits to the data, giving additional freedom to the parameter space. Table~\ref{tab:syserrors-results} shows that their values typically agree with the unity to about $2\%$ for model A and $4$\% for model B. 

Interestingly, the normalization parameters for the data reported by Whitney \etal{}~\cite{Whitney:1974hr} obtained in both models are low but consistent.  It suggests that there might be tension at the level of  $\approx\%$7 between the normalization of this dataset and those of the other measurements, with the results of Ref.~\cite{Whitney:1974hr} being overestimated.

We refer the reader to the Supplemental Material for comparisons of the predictions of both the models to all the measurements considered in this analysis, including the dataset of Whitney \etal{}

\begin{figure*}
\includegraphics[trim={10 0 15 0},clip,width=0.95\textwidth]{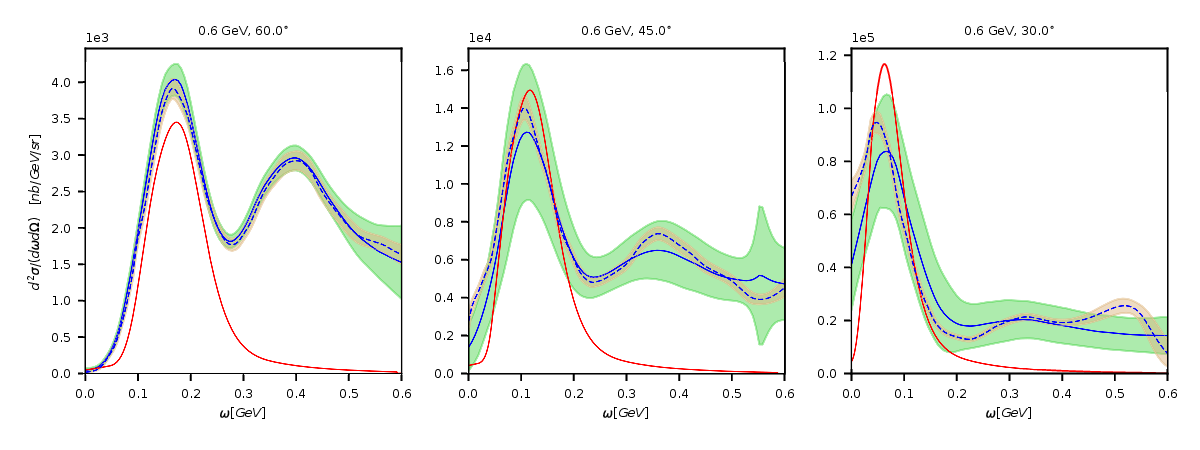}
\caption{Interpolation tests of the obtained fits for beam energy 0.6~GeV and scattering angles 30\degree, 45\degree, and 60\degree. We compare the spectral function calculations~\cite{Ankowski:2014yfa} of the quasielastic cross section, depicted by the solid red lines, with the predictions of model A (model B), represented by the solid (dashed) blue lines. The green (orange) areas correspond to the $1\sigma$ uncertainties.}
\label{Fig:test_clones}
\end{figure*}

So far, we have discussed the performance of the obtained fits in describing the available data. It is interesting to check how well they interpolate the cross section within the domain covered by the training data, and extrapolate beyond it.

To explore the models' abilities to interpolate, in Fig.~\ref{Fig:test_clones}, we compare their predictions with the spectral function calculations of Ref.~\cite{Ankowski:2014yfa}, which include only quasielastic scattering. For these tests, we select the beam energy of 600 MeV---relevant for neutrino-oscillation experiments such as T2K and the Short Baseline Neutrino program---and the scattering angles of 30\degree, 45\degree, and 60\degree. 

The amount of data collected at similar kinematic settings is the largest for $\theta=60\degree$, and the smallest for $\theta=30\degree$. Consequently, the predictions of the two models and their uncertainties are the most similar for $\theta=60\degree$. For lower scattering angles, the uncertainties apparently increase in model A but not in model B. At 45\degree, one can clearly see a sudden jump in the uncertainties of model A at $\omega\simeq0.55$ GeV, reflecting the behavior of the underlying data~\cite{Barreau:1983ht}. 

The predicted positions of the quasielastic peak turn out to agree very well between the predictions of model A and the spectral function approach. This is also the case for model B when  $\theta=60\degree$ and $\theta=45\degree$. However, when we lower the scattering angle, an increasing difference can be observed. 

The magnitude of the cross section predicted by the fits at $\theta=60\degree$ seems consistent with the theoretical model, having in mind a~large contribution of interaction mechanisms other than quasielastic scattering induced by one-body currents. At $\theta=45\degree$, the theoretical calculation agrees with model A (model B) at the $1\sigma$ ($2$--$3\sigma$) level. The differences increase at $30\degree$, showing both the limitations of the obtained fits, and the urgent need for more experimental data to model electroweak interactions with the precision required by the neutrino-oscillation program.

\begin{figure*}
\includegraphics[width=0.65\textwidth]{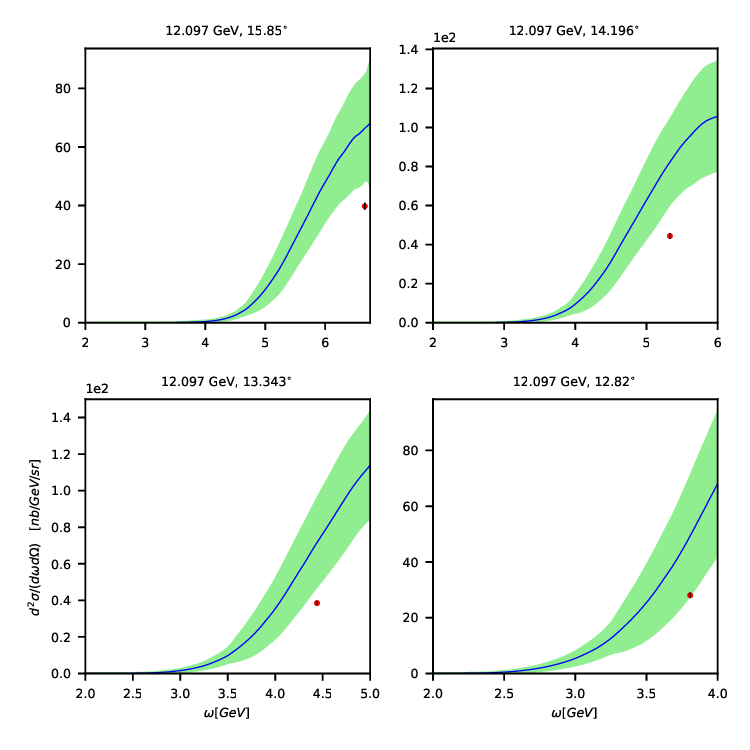}
\caption{Extrapolation tests of model A against the measurements by Gomez \etal~\cite{Gomez:1993ri} in the deep-inelastic regime, not included in the training. The highlighted areas represent the $1\sigma$ uncertainties.}
\label{Fig:Gomez_clone}
\end{figure*}

To test the extrapolation properties of both  models, we compare their predictions for the cross section in the deep-inelastic regime, at the kinematic settings of the measurements by Gomez \etal~\cite{Gomez:1993ri}. As shown in Fig.~\ref{Fig:Gomez_clone}, the predictions of model A manage to reproduce these data at the 1--2$\sigma$ level. For model B (not shown in the picture), we find that it  underestimates  them.

In Fig.~\ref{Fig:Gomez_clone}, the beam energy is fixed to 12.1 GeV, and scattering angles vary between $12.82\degree$ and $15.85\degree$, with the corresponding momentum transfers $4.4\leq\n q\leq7.0$ GeV. Only 2 datasets used in the training probe similar kinematics---the measurements performed by Fomin \etal~\cite{Fomin:2010ei} for beam energy 5.8 GeV, and scattering angles $40\degree$ and $50\degree$, with $3.9\leq\n q\leq5.3$ GeV.

We expect that the accuracy of extrapolation would improve, should more data be available for training.   

In view of the obtained results, we conclude that model A is better at propagating the uncertainties of the underlying experimental data than model~B.

\section{Summary}\label{sec:summary}
In this paper, we develop a~deep neural network framework to parametrize the electron-scattering cross sections for carbon over a broad kinematic region, extending from the quasielastic peak, through resonance excitation, to the onset of deep-inelastic scattering. Our results do not depend on theoretical assumptions but rely exclusively on experimental measurements. We pay special attention to estimating the uncertainties of the network's predictions, which reflect experimental uncertainties and are a~measure of consistency between different datasets. The analysis of the impact of various types (epistemic and aleatoric) of uncertainties on predictive uncertainty is complex \cite{survey_uncertainty_in_dnn} and will be a focus of our future studies.

We discuss two statistical models. Model A is based on an ensemble of $50$ neural networks, which fit bootstrap data{\-}sets. Model B uses a single neural network with drop{\-}out layers. Both approaches reproduce the training data accurately.  Additionally, model A turns out to be able to estimate the cross sections outside the kinematic domain of the training data.   

In addition to their importance for the studies of inclusive electron scattering on nuclei, our empirical fits can be readily included in the {\sc NuWro} Monte Carlo generator and pave the way toward the development of a more accurate description of electroweak cross sections. In future analyses, we plan to adapt and extend the methods developed in this article to the modeling of neutrino interactions.

To facilitate their usage, the resulting fits are released to the public through the GitHub repository~\cite{neuwro}.

\newpage

\begin{acknowledgments}
This research is partly (K.M.G., A.M.A., J.T.S.) or fully (B.E.K., R.D.B., H.P.) supported by the Na{\-}tional Science Centre under grant UMO-2021/41/B/ST2/ 02778.
\end{acknowledgments}

\normalem
\bibliographystyle{apsrev4-2}
\bibliography{bibmoje,bibdrat,bibdata}

\end{document}